\documentclass[11pt]{article}
\usepackage[utf8]{inputenc}
\usepackage{amsfonts}
\usepackage{amsmath}
\usepackage{amssymb}
\usepackage{indentfirst}
\usepackage{graphicx}
\usepackage[colorlinks]{hyperref}
\usepackage{cite}
\usepackage{colortbl}

\usepackage{microtype}
\usepackage[normalem]{ulem}

\numberwithin{equation}{section}
\allowdisplaybreaks

\begin{document}

\baselineskip=18pt 
\baselineskip 0.6cm
\begin{titlepage}
\vskip 4cm

\begin{center}
\textbf{\LARGE Hietarinta Chern-Simons supergravity and its asymptotic structure}
\par\end{center}{\LARGE \par}

\begin{center}
	\vspace{1cm}
    \textbf{Patrick Concha}$^{\dag, \star}$,
     \textbf{Octavio Fierro}$^{\dag, \star}$,
	\textbf{Evelyn Rodríguez}$^{\dag, \star}$
	\small
	\\[5mm]
    $^{\dag}$\textit{Departamento de Matemática y Física Aplicadas, }\\
	\textit{ Universidad Católica de la Santísima Concepción, }\\
\textit{ Alonso de Ribera 2850, Concepción, Chile.}
\\[2mm]
 $^{\star}$\textit{Grupo de Investigación en Física Teórica, GIFT, }\\
	\textit{Concepción, Chile.}
	\\[5mm]
	\footnotesize
        \texttt{patrick.concha@ucsc.cl},
        \texttt{ofierro@ucsc.cl},
	\texttt{erodriguez@ucsc.cl}
	\par\end{center}
\vskip 26pt
\centerline{{\bf Abstract}}
\medskip
\noindent  
In this paper we present the Hietarinta Chern-Simons supergravity theory in three space-time dimensions which extends the simplest Poincaré supergravity theory. After approaching the construction of the action using the Chern-Simons formalism, the analysis of the corresponding asymptotic symmetry algebra is considered. For this purpose, we first propose a consistent set of asymptotic boundary conditions for the aforementioned supergravity theory whose underlying symmetry corresponds to the supersymmetric extension of the Hietarinta algebra. We then show that the corresponding charge algebra contains the super-$\mathfrak{bms}_{3}$ algebra as subalgebra, and has three independent central charges.  We also show that the obtained asymptotic symmetry algebra can alternatively be recovered as a vanishing cosmological constant limit of three copies of the Virasoro algebra, one of which is augmented by supersymmetry.

\end{titlepage}\newpage {\baselineskip=12pt \tableofcontents{}}

\section{Introduction}\label{sec1}
A higher-spin generalization of the D-dimensional Poincaré superalgebra was introduced by Hietarinta in \cite{Hietarinta:1975fu} by including (spinor-)tensor generators associated with (half-)integer higher-spin representations of the Lorentz group. Unlike the infinite-dimensional higher-spin algebras, the (anti-)commutators of higher-spin Hietarinta generators do not close on higher-spin generators. In its simplest form, the Hietarinta algebra defined in three spacetime dimensions contains three spin-2 generators which obey the following structure \cite{Chernyavsky:2020fqs}
\begin{align}
    \left[ J_{a},J_{b}\right] &=\epsilon _{abc}J^{c}\,,& \left[ J_{a},P_{b}\right] &=\epsilon _{abc}P^{c}\,, \notag \\
\left[ J_{a},Z_{b}\right] &=\epsilon _{abc}Z^{c}\,,  & \left[ Z_{a},Z_{b}\right] &=\epsilon _{abc}P^{c}\,. \label{Hiet}
\end{align} 
Here $J_{a}$ are the Lorentz rotation generators, $P_{a}$ correspond to translation generators and $Z_{a}$ is an additional vector generator. The above algebra is also denoted as the Hietarinta/Maxwell algebra \cite{Chernyavsky:2020fqs} due to its isomorphism to the Maxwell algebra \cite{Bacry:1970du,Bacry:1970ye,Schrader:1972zd,Gomis:2017cmt} in which the role of the generators $Z_{a}$ and $P_{a}$ is interchanged. The Hietarinta algebra can be seen as an extension of the Poincaré algebra unlike the Maxwell symmetry, which corresponds to an extension and deformation of Poincaré.

In three spacetime dimensions, Chern-Simons (CS) gravity theories based on both the Maxwell and the Hietarinta  algebras have been largely studied in \cite{Salgado:2014jka,Hoseinzadeh:2014bla,Aviles:2018jzw,Concha:2018zeb,Concha:2021jnn} and \cite{Bansal:2018qyz,Chernyavsky:2019hyp,Chernyavsky:2020fqs}, respectively. Three-dimensional gravity models have received a great interest since they offers us a simple laboratory for studying different aspects of higher-dimensional gravity and the underlying laws of
quantum gravity. On the other hand, the CS theories are characterized by not having local degrees of freedom.  Although the Hietarinta and the Maxwell algebras are isomorphic, they have quiet different physical implications. In particular, the geometries described by the field equations coming from Maxwell gravity are Riemannian (torsionless) and locally flat \cite{Salgado:2014jka,Hoseinzadeh:2014bla,Aviles:2018jzw,Concha:2018zeb,Concha:2021jnn}\footnote{A CS gravity theory based on Maxwell algebra in $2+1$ was initially considered in \cite{Cangemi:1992ri, Duval:2008tr}.}. On the other hand, in the case of Hietarinta gravity theory, the equations of motion describe geometries locally flat but with a non-vanishing torsion. Interestingly, both minimal massive gravity \cite{Bergshoeff:2014pca} and topological massive gravity \cite{Deser:1981wh} appear as particular cases of a more general minimal massive gravity obtained upon spontaneous symmetry breaking in a Hietarinta CS gravity \cite{Chernyavsky:2020fqs}.

To our knowledge, a three-dimensional CS supergravity action based on the simplest Hietarinta superalgebra remains unexplored, despite the Hietarinta superalgebra is known \cite{Bansal:2018qyz}. In the Maxwellian counterpart, there are two known minimal supersymetric extensions of the Maxwell algebra. The first one is called the non-standard Maxwell superalgebra, which includes one fermionic generator \cite{Lukierski:2010dy,Concha:2020atg}, and the other one is denoted as the minimal Maxwell superalgebra allowing for two fermionic generators \cite{Bonanos:2009wy,Bonanos:2010fw,deAzcarraga:2014jpa,Concha:2014xfa,Concha:2014tca,Concha:2015woa,Penafiel:2017wfr,Ravera:2018vra,Concha:2018jxx,Concha:2018ywv,Caroca:2021bjo}.  The non-standard one is not a good candidate to construct a CS supergravity theory since it supersymmetrizes only tensorial generators $Z_{a}$
\begin{align}
    \left\{ Q_{\alpha },Q_{\beta }\right\} &=-\frac{1}{2}\left( C\Gamma
^{a}\right) _{\alpha \beta }Z_{a}\,,
\end{align}
and reproduces an exotic supersymmetric action \cite{Concha:2015woa,Concha:2020atg}. Nonetheless, as it was mentioned in \cite{Chernyavsky:2020fqs}, the translational
generators $P_a$ can be expressed as bilinear expressions of fermionic generators in a supersymmetric extension of the Hietarinta algebra (with one fermionic generator). Thus, unlike the non-standard Maxwell superalgebra, the Hietarinta superalgebra can be seen as the extension of the simple three-dimensional super-Poincaré algebra which strongly suggests that the physical models based on the Hietarinta superalgebra is a priori different from the Maxwell one.

In this work, we explore the construction of a CS supergravity action based on the Hietarinta superalgebra.  The obtained CS action corresponds to an extension of the Poincaré CS supergravity action \cite{Achucarro:1986uwr} whose dynamics is very different to the super-Maxwell one \cite{Concha:2018jxx}. In order to have a better understanding of the additional gauge field  $\sigma^{a}$ related to $Z_{a}$, we study the implications of $\sigma^{a}$ at the level of the asymptotic structure. In asymptotically flat spacetime, the $\mathfrak{bms}_{3}$ algebra, originally formulated by Bondi,  van der Burg, Metzner and Rainer \cite{Bondi:1962px,Sachs:1962zza,Ashtekar:1996cd,Barnich:2010eb}, results to describe the asymptotic symmetry of General Relativity \cite{Barnich:2006av}. The study of richer boundary dynamics could offer a better understanding of the duality beyond the AdS/CFT correspondence \cite{Maldacena:1997re}. Thus, the study of new asymptotic symmetries of CS (super)gravity theories based on symmetries bigger than $\mathfrak{bms}_{3}$ could be worth studying. Such infinite-dimensional algebra have received a growing interest due to recent developments in the derivation for the Weinberg’s soft theorems as well as the
memory effect \cite{Strominger:2013jfa,He:2014laa,Strominger:2014pwa,Avery:2016zce,Hamada:2018vrw,AtulBhatkar:2019vcb,Strominger:2017zoo,Pate:2019mfs}.

Here, we present a novel asymptotic symmetry algebra for three-dimensional flat supergravity. We proposed a set of asymptotic conditions for an extension of the simplest supergravity without cosmological constant.  We show that the corresponding asymptotic symmetry algebra is described in this case by an extension of the usual $\mathfrak{bms}_{3}$ superalgebra \cite{Barnich:2014cwa,Barnich:2015sca,Caroca:2019dds} and can be seen as the supersymmetric extension of the extended l-conformal Galilean algebra for $l=1$ \cite{Chernyavsky:2019hyp}. Moreover, the infinite-dimensional algebra obtained here is characterized by three non-trivial central charges related to the different coupling constants of the CS action and to the CS level $k$. Thus, the asymptotic symmetry superalgebra found here is quiet different to the one appearing in the minimal Maxwell CS supergravity theory, recently introduced in \cite{Matulich:2023xpw}. Indeed, in the Maxwell case, the corresponding asymptotic superalgebra does not contain the $\mathfrak{bms}_{3}$ superalgebra as a subalgebra and possesses two fermionic generators. We also present an alternative way to recover the aforementioned extended super-$\mathfrak{bms}_{3}$ algebra. We show that it appears by applying a vanishing cosmological constant limit $\Lambda\rightarrow 0$ to three copies of the Virasoro algebra, one of which is augmented by supersymmetry, after an appropriate redefinition of the generators is considered. The flat limit at the level of the boundary conditions is also considered.

The paper is organized as follows: In section \ref{sec2} we present the Hietarinta CS supergravity theory in three spacetime dimensions. Section \ref{sec3} is devoted to the study of the corresponding asymptotic symmetry algebra. In section \ref{sec4} we show that the asymptotic symmetry can alternatively be obtained as a vanishing cosmological constant limit of three copies of the Virasoro algebra, one of which is augmented by supersymmetry. Section \ref{concl} is devoted to discussion and possible developments.

\section{Three-dimensional Hietarinta Chern-Simons supergravity}\label{sec2}
In this section, we present the minimal supersymmetric extension of the Hietarinta CS gravity in three dimensions. The underlying symmetry corresponds to the Hietarinta superalgebra \cite{Bansal:2018qyz}, which contains the super-Poincaré symmetry as a subalgebra. The so-called Hietarinta algebra is isomorphic to the three-dimensional Maxwell algebra \cite{Bacry:1970du,Bacry:1970ye,Schrader:1972zd,Gomis:2017cmt}. Although these algebras are isomorphic, they are physically different and leads to distinct CS gravity theories \cite{Chernyavsky:2020fqs}.  

The Hietarinta superalgebra, which we will denote as $s\mathcal{H}$, is spanned by the set of generators $\{J_{a},P_{a},Z_{a},Q_{\alpha}\}$, which satisfy the following (anti)-commutation relations
\begin{align}
    \left[ J_{a},J_{b}\right] &=\epsilon _{abc}J^{c}\,,& \left[ J_{a},P_{b}\right] &=\epsilon _{abc}P^{c}\,, \notag \\
\left[ J_{a},Z_{b}\right] &=\epsilon _{abc}Z^{c}\,,  & \left[ Z_{a},Z_{b}\right] &=\epsilon _{abc}P^{c}\,, \notag \\ 
\left[ J_{a},Q_{\alpha }\right] &=\frac{1}{2}\,\left( \Gamma _{a}\right) _{
\text{ }\alpha }^{\beta }Q_{\beta }\,, & \left\{ Q_{\alpha },Q_{\beta }\right\} &=-\frac{1}{2}\left( C\Gamma
^{a}\right) _{\alpha \beta }P_{a}\,,\label{Hietsu}
\end{align} 
where $a,b,\dots =0,1,2$ are Lorentz indices raised and lowered with the (off-diagonal) Minkowski metric $\eta _{ab}$ and $\epsilon _{abc}$ is the Levi-Civita tensor. The gamma matrices in three dimensions are denoted by $\Gamma_a$ and $C$ is the charge conjugation matrix, satisfying $C^{T}=-C$ and $C\Gamma^{a}=(C\Gamma^{a})^{T}$. 
The CS gravity action based on the Hietarinta algebra was first considered in \cite{Chernyavsky:2020fqs}. Here, we shall extend this construction to the supersymmetric case.

The non-vanishing components of a non-degenerate invariant bilinear product for the Hietarinta superalgebra are given by
\begin{align}
\left\langle J_{a}J_{b}\right\rangle &=\alpha _{0}\eta _{ab}\,,  &  \left\langle J_{a}P_{b}\right\rangle &=\alpha _{1}\eta _{ab}\,, & \left\langle Z_{a}Z_{b}\right\rangle &=\alpha _{1}\eta _{ab}\,,  \notag \\
\left\langle J_{a}Z_{b}\right\rangle &=\alpha _{2}\eta _{ab}\,, & 
\left\langle Q_{\alpha }Q_{\beta }\right\rangle &=\alpha _{1}C_{\alpha \beta
}\,, \label{ITHie}
\end{align}
where $\alpha_0$, $\alpha_1$ and $\alpha_2$ are arbitrary constants.
The gauge connection one-form $A$ reads
\begin{equation}
A=\omega ^{a}J_{a}+e^{a}P_{a}+\sigma ^{a}Z_{a}+\bar{\psi}Q
\,,  \label{1fP}
\end{equation}%
where $\omega^{a}$ is the spin connection one-form, $e^{a}$ corresponds to
the dreibein one-form, $\sigma ^{a}$ is the gauge field one-form associated with the vector generator $Z_a$,
while $\psi $ is a fermionic gauge field.
The corresponding curvature two-form $F=dA+\frac{1}{2}[A,A]$ is given by
\begin{equation}
F=R^{a}J_{a}+\mathcal{T}^{a}P_{a}+\mathcal{F}^{a}Z_{a}+\nabla \bar{\psi}
Q \,,
\end{equation}
with
\begin{eqnarray}
R^{a} &=&d\omega ^{a}+\frac{1}{2}\epsilon ^{abc}\omega _{b}\omega _{c}\,,
\notag \\
\mathcal{T}^{a} &=&de^{a}+\epsilon ^{abc}\omega _{b}e_{c}+\frac{1}{2}\epsilon _{\text{ }}^{abc}\sigma_{b}\sigma_{c}+\frac{1}{4} i \bar{\psi}\Gamma ^{a}\psi \,,  \notag \\
\mathcal{F}^{a} &=&d\sigma ^{a}+\epsilon ^{abc}\omega _{b}\sigma _{c}
\,. \label{bosc}
\end{eqnarray}
The covariant derivative $\nabla =d+[A,\cdot ]$ acting on spinors reads
\begin{equation}
\nabla \psi=d\psi +\frac{1}{2}\,\omega ^{a}\Gamma _{a}\psi \,.   \label{ferc}
\end{equation}
The CS supergravity action invariant under the Hietarinta superalgebra is obtained considering the non-vanishing components of the invariant tensor (\ref{ITHie}) and the gauge
connection one-form (\ref{1fP}) in the general expression of a CS action
\begin{equation}
I[A]=\frac{k}{4\pi }\int_{\mathcal{M}}\left\langle AdA+\frac{2}{3}
A^{3}\right\rangle \,,  \label{CSaction}
\end{equation}
defined on a three-dimensional manifold $\mathcal{M}$, and where $k=\frac{1}{4G}$ is the CS level of the theory related to the gravitational constant $G$. 

Then, the Hietarinta CS supergravity action reads
\begin{eqnarray}
I_{s\mathcal{H}} &=&\frac{k}{4\pi }\int \alpha _{0}\left( \,\omega
^{a}d\omega _{a}+\frac{1}{3}\,\epsilon _{abc}\omega ^{a}\omega ^{b}\omega
^{c}\right)  \notag \\
&&+\alpha _{1}\left( 2e^{a}R_{a}+\sigma^{a}\mathcal{F}_{a} -i\bar{\psi}\nabla \psi \right) \,+2\alpha
_{2} \sigma ^{a} R_{a} \,,  \label{sHCS}
\end{eqnarray}
which can be seen as an extension of the three-dimensional Poincaré CS supergravity action \cite{Achucarro:1987vz} without introducing a cosmological constant. From the previous action we can see that the term proportional to $\alpha _{0}$ contains the gravitational CS 
Lagrangian \cite{Witten:1988hc,Zanelli:2005sa}. The term along $\alpha _{1}$ contains the Einstein-Hilbert term, a term involving  the $\sigma ^{a}$ field and the
Rarita-Schwinger Lagrangian, while $\alpha _{2}$ yields a term involving the field $\sigma ^{a}$. Naturally, the three-dimensional Poincaré CS supergravity action \cite{Achucarro:1987vz} is recovered when the additional gauge field $\sigma_a$ is switched off. As we shall see, the presence of the $\sigma_a$ gauge field has implications in the asymptotic structure of the theory.  In absence of supersymmetry, the CS action corresponds to the Hietarinta CS gravity, in which the role of the vielbein $e^{a}$ (associated with the Poincaré translations)
and of the additional spin-2 field $\sigma^{a}$ get interchanged in comparison to the Maxwell CS gravity action \cite{Salgado:2014jka,Hoseinzadeh:2014bla,Aviles:2018jzw,Concha:2018zeb}. 

The equations of motion are obtained by the extremization of the action, which gives
\begin{eqnarray}
\delta e^{a} & : & \qquad0=\alpha_{1}R_{a}\,,\nonumber \\
\delta\omega^{a} & : & \qquad0=\alpha_{0}R_{a}+\alpha_{1}\left(T_{a}+\frac{1}{2}\epsilon_{abc}\sigma^{b}\sigma^{c}+\frac{1}{4} i\bar{\psi}\Gamma_{a}\psi \right)+\alpha_{2}\mathcal{F}_{a}\,, \nonumber\\
\delta\sigma^{a} & : & \qquad0=\alpha_{1}\mathcal{F}_{a}+\alpha_{2}R_{a}\,,\nonumber \\
\delta\bar\psi & : & \qquad0=\alpha_{1}\nabla\psi\,, \label{cs5-1}
\end{eqnarray}
where $T^a =De^a$ is the usual torsion two-form. Let us note that the non-degeneracy of the invariant bilinear trace is guaranteed for $\alpha_{1}\neq0$ which implies that the above equations can be equivalently written as
\begin{equation}
\mathcal{T}^{a} =0\,, \qquad R^{a} =0\,,\qquad \mathcal{F}^{a}=0\,,\qquad \nabla\psi\,.\label{eom1}
\end{equation}
Thus, the field equations derived from the CS supergravity action \eqref{sHCS} reduce to the vanishing of the
curvature two-forms \eqref{bosc}-\eqref{ferc}. The CS action (\ref{sHCS}) is invariant, by construction, under the gauge
transformation $\delta A=d\Lambda +\left[ A,\Lambda \right] $, with gauge parameter $\Lambda =\chi ^{a}J_{a}+\varepsilon
^{a}P_{a}+\gamma ^{a}Z_{a}+\bar{\epsilon}Q$. In
particular, the action is invariant under the following local supersymmetry
transformation laws
\begin{eqnarray}
\delta \omega ^{a} &=&0\,,  \notag \\
\delta e^{a} &=&\frac{1}{2}\,i\bar{\epsilon}\Gamma ^{a}\psi \,,  \notag \\
\delta \sigma ^{a} &=&0 \nonumber \\
\delta \psi &=&d\epsilon +\frac{1}{2}\,\omega ^{a}\Gamma _{a}\epsilon \,.  \label{Hsusy}
\end{eqnarray}
One can notice that curvature two-forms \eqref{bosc}-\eqref{ferc} transform covariantly under the supersymmetry transformation laws \eqref{Hsusy}.

\subsection{Bosonic solutions in the BMS gauge}

In this section we analyze the bosonic field equations (\ref{eom1}) (with $\psi=0$). We consider spacetimes with null boundary, which can be described in the BMS gauge.  We parametrize spacetime by the local coordinates $x^{\mu}=\left(u,r,\phi\right)$, where $-\infty < u <\infty$ is the retarded time coordinate,  $\phi \sim \phi + 2 \pi$ is the angular coordinate and the boundary is located at $r=\it{const}$. Then, the metric can be written as follows \cite{Barnich:2014cwa}
\begin{equation}
    ds^{2}=\mathcal{M}du^{2}-2dud r +\mathcal{N}d\phi d u+r^{2}d \phi^{2}\,.
\end{equation}
where $\mathcal{M}$ and $\mathcal{N}$ are two arbitrary functions of the coordinates $u,\phi$. As was previously mentioned, the Hietarinta symmetry can be obtained from the Maxwell one, in which the role of the generators $P_a$ and $Z_a$ gets interchanged ($P_a\leftrightarrow Z_a$). Then, one way of finding the solutions of the Hietarinta gravity theory is considering the solutions of the Maxwell theory found in \cite{Concha:2018zeb}, and then go to the basis we are interested in, by interchanging the role of the vielbein and the gauge field along $Z$. 

In the Maxwell case, the fields $(\Tilde{\omega}^a,\Tilde{e}^a, \Tilde{\sigma}^a$) obey boundary conditions in which the functions $\mathcal{M}$ and $\mathcal{N}$ are given for the known results in asymptotically flat gravity in three dimensions
\begin{equation}\label{M,N}
    \mathcal{M}=\mathcal{M}(\phi)\,,\qquad \mathcal{N}=\mathcal{J}(\phi)+u \mathcal{M}^{\prime}(\phi)\,.
\end{equation}
 The spacetime line element can be written in terms of the vielbein as $ds^2 = \eta_{ab} \Tilde{e}^{a} \Tilde{e}^{b}$, where $\eta_{ab}$ is the off-diagonal Minkowski metric. Then, in the Maxwell case, the vielbein and the torsionless spin connection one-forms are given by
\begin{align}
     \Tilde{e}^{0}&=-dr+\frac{1}{2}\mathcal{M}du+\frac{1}{2}\mathcal{N}d\phi\,, & \Tilde{e}^{1}&= d u\,, & \Tilde{e}^{2}&= rd\phi \,, \nonumber \\ 
     \Tilde{\omega}^{0}&=\frac{1}{2}\mathcal{M}d\phi\,, & \Tilde{\omega}^{1}&= d\phi\,, & \Tilde{\omega}^{2}&=0\,.
\end{align}
Furthermore, solving the e.o.m involving the gauge field $\Tilde{\sigma}^a$, it was shown in \cite{Concha:2018zeb} that it can be written in the following way 
\begin{equation}
    \Tilde{\sigma}^{0}=\frac{1}{2}\mathcal{N} du+\frac{1}{2}\left(\mathcal{F}-r^{2}\right)d\phi , \qquad\Tilde{\sigma}^{1}= 0, \qquad\Tilde{\sigma}^{2}= 0\,,
\end{equation}
where $\mathcal{F}=\mathcal{F}(u,\phi)$ is given by
\begin{equation}
    \mathcal{F}=\mathcal{Z}(\phi)+u\mathcal{M}^{\prime}(\phi)+\frac{u^{2}}{2}\mathcal{J}^{\prime\prime}(\phi)\,.
\end{equation}

Here, we are interested in the Hietarianta symmetry which appears from the Maxwell algebra when the generators $P_a$ and $Z_a$ are interchanged, or equivalently, when the role of the vielbein and the Maxwell gauge field is exchanged. Then, the Hietarinta gauge fields $\left({e}^a, {\omega}^a, {\sigma}^a\right)$ are related to the Maxwell ones $(\Tilde{e}^a, \Tilde{\omega}^a, \Tilde{\sigma}^a)$ as follows: 
\begin{equation}
    e^a=\Tilde{\sigma}^a\,,\qquad \omega^a=\Tilde{\omega}^a\,,\qquad \sigma^a=\Tilde{e}^a\,.
    \end{equation}
Consequently, the bosonic counterpart of the field equations (\ref{eom1}) (with $\psi=0$) are solved by the following components of the gauge fields
\begin{align}
    e ^{0} & =\frac{1}{2}\mathcal{N} du+\frac{1}{2}\left(\mathcal{F}-r^{2}\right)d\phi\,,  &
\omega ^{0} & =\frac{1}{2}\mathcal{M}d\phi\,, & \sigma ^{0} & =-dr+\frac{1}{2}\mathcal{M}du+\frac{1}{2}\mathcal{N}d\phi\,,\nonumber \\ 
e ^{1} & =0\,, & \omega ^{1} & =d\phi\,, & \sigma ^{1} & =du\,,  \nonumber \\
e ^{2} & =0\,, & \omega ^{2} & =0\,, & \sigma^{2} &
=r d\phi\,. \label{ewf}
\end{align}

\section{Asymptotic symmetry algebra}\label{sec3}
In this section, we compute the asymptotic symmetry algebra for the previously presented Hietarinta CS supergravity. To this end we provide the suitable fall-off
conditions for the gauge fields at infinity and the gauge transformations preserving the boundary
conditions. Let us mention that the  boundary conditions we will consider here correspond to a supersymmetric extension of the conditions presented in \cite{Concha:2018zeb}, but considering the exchange of  the gauge fields $e^{a}$ and $\sigma^{a}$. Then, the charge algebra is found using the Regge-Teitelboim method \cite{Regge:1974zd}. 

\subsection{Boundary conditions}

Based on the results obtained in the previous section, we consider the following behaviour of the gauge fields at the boundary 
\begin{eqnarray}
A & = & \frac{1}{2}\mathcal{M}d\text{\ensuremath{\phi}}J_{0}+d\phi J_{1}+\frac{1}{2}\left(\mathcal{N}du+\mathcal{F}d\phi-r^{2}d\phi\right)P_{0}+rdu P_{2}\nonumber \\
 &  & +\left(-dr+\frac{1}{2}{\cal M}du+\frac{1}{2}\mathcal{N}d\phi\right)Z_{0}+du Z_{1}+rd\phi Z_{2}+\frac{\psi}{2^{1/4}}d\phi\, Q_{+}\,,\label{AsM}
\end{eqnarray}
where the functions ${\cal M}$, ${\cal N}$ and $\cal{F}$, and the Grassmann-valued spinor component $\psi$ are assume to depend on all boundary coordinates $x^{i}=(u,\phi)$.
The radial dependence of the gauge field $A$ can be dropped out by the gauge transformation 
\begin{equation}\label{gtransformation}
A=h^{-1}dh+h^{-1}ah\,,
\end{equation}
where $h=$ e$^{-rZ_{0}}$. Then, the new gauge field $a=a_{u}du+a_{\phi}d\phi$ becomes the asymptotic field, where the component along $\phi$ is
\begin{eqnarray}
a_{\phi} &= & \frac{1}{2}{\cal  M}d\phi\, J_{0}+d\phi\, J_{1}+\frac{1}{2}\mathcal{F}d\phi\,P_{0}+\frac{1}{2}{\cal N}d\phi Z_{0} +\frac{\psi}{2^{1/4}}d\phi\, Q_{+}\,.\label{aa}
\end{eqnarray}
The asymptotic symmetries correspond to the set of transformations 
that leaves the asymptotic conditions (\ref{AsM}) invariant. Thus, we consider gauge parameters of the form $\Lambda=h^{-1}\lambda h$ and 
\begin{align}
\lambda = &\,\chi^{a}(u,\phi)\,J_{a}+\varepsilon^{a}(u,\phi) P_{a}+\gamma^{a}(u,\phi) Z_{a}+2^{1/4}\epsilon^{+}(u,\phi)Q_{+}+2^{1/4}\epsilon^{-}(u,\phi)Q_{-}\,.\label{lambda}
\end{align}
Considering the gauge connection \eqref{aa} and the gauge parameter \eqref{lambda}, it is possible to show that the parameters can be solved in terms of three arbitrary bosonic functions and one arbitrary fermionic function, i.e $\lambda=\lambda(Y,R,T,\mathcal{E})$. Indeed, we find
\begin{align}
   \chi^{0} & =\frac{\mathcal{M}}{2}\,Y-Y^{\prime\prime}\ &  \varepsilon^{0} &=\frac{1}{2}\left(\mathcal{M}R+\mathcal{N}T+\mathcal{F} Y\right)+\frac{1}{2}\epsilon^{-}\psi-R^{\prime\prime}\ &\gamma^{0} & =\frac{1}{2}\left(\mathcal{M}T+\mathcal{N}Y\right)-T^{\prime\prime}, \nonumber\\
   \chi^{2} & =-Y^{\prime}\,, & \varepsilon^{2} & =-R^{\prime}\,, & \gamma^{2} & =-T^{\prime}\,, \nonumber \\
   \chi^{1} & =Y\,, & \varepsilon^{1} & =R\,, & \gamma^{1} & =T\,,
\end{align}
while the fermionic functions read
\begin{equation}
    \epsilon^{-}=\mathcal{E}\,, \qquad \epsilon^{+}=\frac{1}{\sqrt{2}}\left(Y\psi-2\mathcal{E}^{\prime}\right)\,.
\end{equation}
Then, the transformation laws for the arbitrary functions $\mathcal{M}$, $\mathcal{N}$, $\mathcal{F}$ and $\Psi$ under the asymptotic symmetries are given by
\begin{eqnarray}
\delta{\cal M} & = & {\cal M}^{\prime}Y+2{\cal M}Y^{\prime}-2Y{}^{\prime\prime\prime}\quad\,,\nonumber \\
\delta{\cal N} & = & {\cal N}^{\prime}Y+2{\cal N}Y^{\prime}+{\cal M}^{\prime}T+2{\cal M}T^{\prime}-2T{}^{\prime\prime\prime}\,,\nonumber\\
\delta{\cal F} & = & {\cal F}^{\prime}Y+2{\cal F}Y^{\prime}+{\cal N}^{\prime}T+2{\cal N}T^{\prime}+{\cal M}^{\prime}R+2{\cal M}R^{\prime}+3 i\mathcal{E^{\prime}}\Psi+i \mathcal{E}\Psi^{\prime}-2R{}^{\prime\prime\prime}\,,\nonumber \\
\delta{ \Psi} & = &\Psi^{\prime}Y+\frac{3}{2}\Psi Y^{\prime}+\frac{1}{2}\mathcal{M}\mathcal{E}-2\mathcal{E}^{\prime\prime}\,. \label{transflaw}
\end{eqnarray}
Let us now determine the asymptotic form of the gauge fields along time evolution. For this purpose we incorporate the Lagrange multipliers for every dynamical field in the asymptotic form of the gauge field.  The asymptotic symmetries along time will be preserved whenever the Lagrange multiplier is $A_{u}=h^{-1}a_{u}h$, with 
\begin{equation}
    a_{u}=\lambda(\mu_{\mathcal{N}},\mu_{\mathcal{M}},\mu_{\mathcal{F}},\mu_{\Psi})\,,
\end{equation}
where the "chemical potentials" are arbitrary functions and assumed to be fixed at the boundary. The time  evolution of the gauge fields in the asymptotic region is is given by the following conditions
\begin{eqnarray}
\dot{\cal M} & = & {\cal M}^{\prime}\mu_{\mathcal{N}}+2{\cal M}\mu_{\mathcal{N}}^{\prime}-2\mu{}^{\prime\prime\prime}_{\mathcal{N}}\quad\,,\nonumber \\
\dot{\cal N} & = & {\cal N}^{\prime}\mu_{\mathcal{N}}+2{\cal N}\mu_{\mathcal{N}}^{\prime}+{\cal M}^{\prime}\mu_{\mathcal{F}}+2{\cal M}\mu_{\mathcal{F}}^{\prime}-2\mu{}^{\prime\prime\prime}_{\mathcal{F}}\,,\nonumber\\
\dot{\cal F} & = & {\cal F}^{\prime}\mu_{\mathcal{N}}+2{\cal F}\mu_{\mathcal{N}}^{\prime}+{\cal N}^{\prime}\mu_{\mathcal{F}}+2{\cal N}\mu_{\mathcal{F}}^{\prime}+{\cal M}^{\prime}\mu_{\mathcal{M}}+2{\cal M}\mu_{\mathcal{M}}^{\prime}+3 i\mu_{\Psi}^{\prime}\Psi+i \mu_{\Psi}\Psi^{\prime}-2\mu{}^{\prime\prime\prime}_{\mathcal{M}}\,,\nonumber \\
\dot{ \Psi} & = &\Psi^{\prime}\mu_{\mathcal{N}}+\frac{3}{2}\Psi \mu_{\mathcal{N}}^{\prime}+\frac{1}{2}\mathcal{M}\mu_{\Psi}-2\mu_{\Psi}^{\prime\prime}\,. \label{transflaw}
\end{eqnarray}

The above transformation laws and conditions for the asymptotic field contain the information of the asymptotic structure of the Hietarinta CS supergravity and their corresponding algebra. Indeed, the charge algebra of the Hietarinta supergravity theory can be computed following the Regge-Teitelboim approach \cite{Regge:1974zd}. In what follows we will consider this construction.

\subsection{Charge algebra: Extended super-$\mathfrak{bms}_{3}$ algebra }
The charge algebra of the Hietarinta CS supergravity theory in representation of Poisson brackets can be obtained using the Regge-Teitelboim method \cite{Regge:1974zd} directly from the transformation law
\begin{equation}\label{rt}
\delta_{\Lambda_{2}}Q[\Lambda_{1}]=\left\{ Q[\Lambda_{1}],Q[\Lambda_{2}]\right\} ,
\end{equation}
where $Q[\Lambda]$ is the conserved charges spanning the algebra \cite{Banados:1994tn}. On the other hand, the variation of the charge in CS theory is given by 
\begin{equation}
\delta Q[\Lambda]=\frac{k}{2\pi}\int\limits _{\partial\Sigma}\left\langle \Lambda\delta A\right\rangle \,.
\end{equation}
After applying the gauge transformation \eqref{gtransformation} which introduces the asymptotic field (\ref{aa}) we get \cite{Banados:1998gg}
\begin{equation}
\delta Q[\lambda]=\frac{k}{2\pi}\int d\phi\left\langle \lambda\delta a_{\phi}\right\rangle \,. \label{QQ}
\end{equation}
Considering the invariant tensor \eqref{ITHie} and the gauge field $a$ defined in \eqref{aa} in the previous expression, we obtain
\begin{equation}\label{charge-variation-01}
    \delta Q [Y,R,T, \mathcal{E} ]= \int d\phi \left( Y \delta \mathbf{J}+ R\delta \mathbf{P} + T\delta \mathbf{Z}+2i\mathcal{E} \delta \mathbf{\Psi} \right)
\end{equation}
where we have defined
\begin{align}
   \mathbf{J} &= \frac{k}{4\pi}\left(\alpha_2\mathcal{N}+\alpha_0\mathcal{M}+\alpha_1\mathcal{F}\right)\, , \\
   \mathbf{P}&=  \frac{k}{4\pi}\alpha_{1}\mathcal{M} \,, \\
   \mathbf{Z}&= \frac{k}{4\pi}\left( \alpha_{1} \mathcal{N}+\alpha_{2}\mathcal{M}\right) \,, \\
   \mathbf{\Psi} &=-\frac{k}{4\pi}\alpha_{1}\Psi\,.
\end{align}

We assume that the functions $Y$, $T$, $R$ and $\mathcal{E}$ do not depend on the fields which implies that the charge variation is integrable on the phase space. Then, one finds
\begin{align}
 Q [Y,R,T, \mathcal{E} ]= \int d\phi \left( Y \mathbf{J}+ R \mathbf{P} + T \mathbf{Z}+2i\mathcal{E} \mathbf{\Psi} \right)\,.\label{QQ2}
\end{align}
There are four independent surface charges,
\begin{equation}
    j[Y] = Q[Y,0,0,0] \, , \qquad p[R] = Q[0,R,0,0] \, , \qquad z[T] = Q[0,0,T,0]  \, , \qquad g[\mathcal{E}]=Q[0,0,0,\mathcal{E}]\,,
\end{equation}
associated with four independent symmetry generators $Y$, $R$ and $T$ and $\mathcal{E}$. Then, the Poisson brackets of these independent charges can be evaluated considering \eqref{rt} and \eqref{transflaw}. Expanding in Fourier modes and defining
\begin{equation}
{\cal J}_{m}=j[e^{im\text{\ensuremath{\phi}}}]\text{\,,\qquad}{\cal P}_{m}=p[e^{im\text{\ensuremath{\phi}}}]\text{\,,\qquad}{\cal Z}_{m}=z[e^{im\text{\ensuremath{\phi}}}]\text{\,,\qquad}{\cal G}_{m}=g[e^{im\text{\ensuremath{\phi}}}]\,,
\end{equation}
one obtain the following Poisson brackets
\begin{align}
 i\left\{ \mathcal{J}_{m},\mathcal{J}_{n}\right\} & =  \left(m-n\right)\mathcal{J}_{m+n}+\dfrac{c_{1}}{12}\,m^{3}\delta_{m+n,0}\,,\nonumber\\
i\left\{ \mathcal{J}_{m},\mathcal{P}_{n}\right\}  & =  \left(m-n\right)\mathcal{P}_{m+n}+\dfrac{c_{2}}{12}\,m^{3}\delta_{m+n,0}\,,\nonumber\\
i\left\{ \mathcal{P}_{m},\mathcal{P}_{n}\right\} & = 0\,,\nonumber\\
i\left\{ \mathcal{Z}_{m},\mathcal{Z}_{n}\right\}  & =  \left(m-n\right)\mathcal{P}_{m+n}+\dfrac{c_{2}}{12}\,m^{3}\delta_{m+n,0}\,,\nonumber\\
i\left\{ \mathcal{J}_{m},\mathcal{Z}_{n}\right\}  & =  \left(m-n\right)\mathcal{Z}_{m+n}+\dfrac{c_{3}}{12}\,m^{3}\delta_{m+n,0}\,,\nonumber\\
i\left\{ \mathcal{P}_{m},\mathcal{Z}_{n}\right\}  & = 0
\,,\nonumber\\
i\left\{ \mathcal{J}_{m},\mathcal{G}_{n}\right\}  & =  \left(\frac{m}{2}-n\right)\mathcal{G}_{m+n}\,,\nonumber\\
i\left\{ \mathcal{P}_{m},\mathcal{G}_{n}\right\}  & = 0\,,\nonumber\\
i\left\{ \mathcal{G}_{m},\mathcal{G}_{n}\right\}  & = \mathcal{P}_{m+n}+\dfrac{c_{2}}{6}\,m^{2}\delta_{m+n,0}\,, \label{asymsH}
\end{align}
where we have used the integral representation of the Kronecker delta $\delta_{m,n}=\frac{1}{2\pi}\int d\phi\,e^{i(m-n)\text{\ensuremath{\phi}}}$. As we can see from the previous algebra it defines an extension of the super-$\mathfrak{bms}_{3}$ algebra spanned by $\mathcal{J}_{m}$, $\mathcal{P}_{m}$ and $\mathcal{G}_{m}$ and corresponds to a supersymmetric extension of the extended l-conformal Galiean algebra when $l=1$ \cite{Chernyavsky:2019hyp}. Let us note that the present superalgebra can be seen as a non-trivial central extension since the constant $\frac{c_i}{12}m^{3}\delta_{m+n,0}$ cannot be removed by redefining the generators $\mathcal{J}_m$, $\mathcal{P}_m$ and $\mathcal{Z}_m$. As it is expected, it corresponds to an infinite-dimensional lift of the Hietarinta superalgebra, with three central charges which are related to the CS level $k$ and to the three coupling constants appearing in the CS supergravity action through the following relation 
\begin{equation}
    c_{i}=12 k\alpha_{i-1}\,, \qquad i=1,2,3\,.
\end{equation}
Indeed, the Hietarinta superalgebra is a finite subalgebra of (\ref{asymsH}). It can be explicitly seen by identifying the modes in (\ref{asymsH}) with the generators in \eqref{Hietsu} as follows
\begin{align}
    \mathcal{J}_{-1}&=-\sqrt{2}J_{0}\,, & \mathcal{J}_{1}&=\sqrt{2}J_{1}\,, &   \mathcal{J}_{0}&=J_{2}\,, \nonumber
    \\
    \mathcal{P}_{-1}&=-\sqrt{2}P_{0}\,, & \mathcal{P}_{1}&=\sqrt{2}P_{1}\,, &   \mathcal{P}_{0}&=P_{2}\,, \nonumber \\
\mathcal{Z}_{-1}&=-\sqrt{2}Z_{0}\,, & \mathcal{Z}_{1}&=\sqrt{2}Z_{1}\,, &   \mathcal{Z}_{0}&=Z_{2}\,, \nonumber\\
  \mathcal{G}_{-1/2}&=\sqrt{2}Q_{+}\,, & \mathcal{G}_{1/2}&=\sqrt{2}Q_{-}\,.
\end{align}
Unlike the super-$\mathfrak{bms}_3$ algebra, the obtained asymptotic superalgebra contains an additional central charge $c_3=12k\alpha_2$. Naturally, the centrally extended super-BMS3 algebra\cite{Barnich:2014cwa,Barnich:2015sca,Caroca:2019dds} is recovered when $\mathcal{Z}_m$ and $c_3$ are switched off.

Let us mention that when the role of the generators
$\mathcal{Z}_{m}$ and $\mathcal{P}_{m}$ is interchanged,  the charge algebra $\eqref{asymsH}$ corresponds to the supersymmetric extension of the asymptotic symmetry algebra of the Maxwell CS gravity, given by a deformation of the $\mathfrak{bms}_{3}$ algebra \cite{Concha:2018zeb,Caroca:2017onr}. Nonetheless, it is important to clarify that the exchange of $\mathcal{Z}_{m}$ and $\mathcal{P}_{m}$ reproduces an infinite-dimensional lift of the non-standard Maxwell superalgebra \cite{Lukierski:2010dy,Concha:2020atg}. The corresponding asymptotic symmetry algebra of the minimal Maxwell CS supergravity has been recently presented in \cite{Matulich:2023xpw}, which is quiet different to the asymptotic algebra introduced here. In the minimal Maxwell case, the asymptotic algebra is characterized by two fermionic generators $\mathcal{G}_m$ and $\mathcal{H}_m$.
\section{Extended super-$\mathfrak{bms}_{3}$ algebra and vanishing cosmological constant limit} \label{sec4}

Let us note that the infinite-dimensional algebra (\ref{asymsH}) can be obtained as a Inönü-Wigner contraction of the direct sum of three copies of the Virasoro algebra, one of which augmented by supersymmetry. Indeed, let us consider the the direct product $\mathfrak{svir}\oplus\mathfrak{vir}\oplus\mathfrak{vir}$ as follows 
\begin{equation}
\begin{array}{lcl}
i\left\{ \mathcal{L}_{m}^{\pm},\mathcal{L}_{n}^{\pm}\right\}  & = & \left(
m-n\right) \mathcal{L}_{m+n}^{\pm}+\dfrac{c^{\pm}}{12}\,m^3
\delta _{m+n,0}\,, \\[5pt]
i\left\{ \hat{\mathcal{L}}_{m},\hat{\mathcal{L}}_{n}\right\}  & = & \left(
m-n\right) \mathcal{\hat{L}}_{m+n}+\dfrac{\hat{c}}{12}\,m^3\delta _{m+n,0}\,, \\ [5pt]
i\left\{\mathcal{L}_{m}^{+},\mathcal{Q}_{r}^{+}\right\} & = & \left( \dfrac{m}{2}%
-r\right) \mathcal{Q}_{m+r}^{+}\,\,, \\[5pt]
i\left\{ \mathcal{Q}_{r}^{+},\mathcal{Q}_{s}^{+}\right\} & = & \mathcal{L}_{r+s}^{+}+
\dfrac{c^{+}}{6}\, r^{2} \delta _{r+s,0}\,.
\end{array}
\end{equation}
After the following redefinitions,
\begin{align}
  \mathcal{L}_{m}^{+}  &=\frac{1}{2}\left(\ell^{2}\mathcal{P}_m +\ell \mathcal{Z}_m\right)\,, & \mathcal{L}_{m}^{-}  &=\frac{1}{2}\left(\ell^{2}\mathcal{P}_{-m} -\ell \mathcal{Z}_{-m}\right)\,,& \hat{\mathcal{L}}_{m}=\mathcal{J}_{-m}-\ell^{2}\mathcal{P}_{-m}\,, \notag \\
Q_{r}^{+}&=\frac{\ell}{\sqrt{2}}\mathcal{G}_{r}\,,&
  c^{\pm}&=\frac{1}{2}\left(\ell^{2}c_2\pm\ell c_3\right)\,, & \hat{c} =\left(c_1-\ell^{2}c_2\right)\,.
\end{align}
the previous algebra, written in the basis $\lbrace \mathcal{J}, \mathcal{P}, \mathcal{Z},\mathcal{G}\rbrace$, leads to the extension of super-$\mathfrak{bms}_{3}$ algebra given in \eqref{asymsH} in the limit $\ell\rightarrow\infty$. Let us notice that the basis $\lbrace \mathcal{J}, \mathcal{P}, \mathcal{Z},\mathcal{G}\rbrace$ satisfy a supersymmetric extension of the infinite-dimensional lift of the AdS-Lorentz algebra \cite{Caroca:2019dds} in which the role of $\mathcal{P}$ and $\mathcal{Z}$ is interchanged. In such basis, the $\ell$ parameter is related to the cosmological constant through $\Lambda=-\frac{1}{\ell^2}$. Then, the limit $\ell\rightarrow\infty$ can be seen as a vanishing cosmological constant limit. It is important to mention that the obtained asymptotic symmetry algebra along the one of the minimal Maxwell supergravity \cite{Matulich:2023xpw} can both be obtained as a flat limit $\ell\rightarrow\infty$. Nonetheless, the asymptotic algebra of the minimal Maxwell supergravity appear by contracting three copies of the Virasoro algebra, two of which are augmented by supersymmetry \cite{Caroca:2019dds,Matulich:2023xpw}. 

It is also possible to show that the boundary conditions \eqref{aa} can be derived as a flat limit of a supersymmetric extension of the  Brown-Henneaux boundary conditions considered in \cite{Concha:2018jjj}, after an appropriate change of basis.  First, let us consider the basis $\left\lbrace J^{\pm}_{a},\hat{J}_{a}\right\rbrace$, in which the connection one-form can be written as the sum of three $\mathfrak{sl}(2,\mathbb{R})$ connections: $A=A^{+}+A^{-}+\hat{A}$. The CS action splits into the sum of three Lorentz CS actions, one for each connection, leading to three $\mathfrak{sl}(2,\mathbb{R})$ connections satisfying Brown-Henneaux boundary conditions \cite{Concha:2018jjj}:
\begin{equation}
    a^{\pm}=\left(\mathcal{L}^{\pm}J_{0}^{\pm}+J_{1}^{\pm}\right)dx^{\pm}\,, \qquad \hat{a}=\left(\mathcal{L}\hat{J}_{0}+\hat{J}_{1}\right)d\phi\,,
\end{equation}
where $x^{\pm}=\phi\pm\frac{t}{\ell}$ and 
\begin{equation}
    \mathcal{L}^{\pm}=\mathcal{L}^{\pm}\left(x^\pm\right)\,, \qquad \mathcal{L}=\mathcal{L}\left(\phi\right)\,,
\end{equation}
are arbitrary functions of their arguments, which are required to satisfy on-shell,
\begin{equation}
    \partial_{\mp}\mathcal{L}^{\pm}=0\,, \qquad \partial_{u}\mathcal{L}=0\,.\label{chiral1}
\end{equation}
Thus, each set of charges resulting from \eqref{QQ} will satisfy a Virasoro algebra with central charges $c^{\pm}$ and $\hat{c}$. 

Then, let us consider three copies of the $\mathfrak{sl}(2,\mathbb{R})$ algebra, one of them augmented by supersymmetry, and where each copy is spanned by generators $\left\lbrace J^{+}_{a}, Q_{\alpha}\right\rbrace$, $\left\lbrace J^{-}_{a}\right\rbrace$, $\left\lbrace\hat{J}_{a}\right\rbrace$. The $r$-independent gauge connection will be written in terms of a supersymmetric extension of the Brown-Henneaux  boundary conditions:
\begin{align}
    a^{+}&=\left(\mathcal{L}^{+}J_{0}^{+}+J_{1}^{+}+\frac{\psi}{2^{1/4}}\, Q_{+}\right)dx^{+}\,, \nonumber \\
    a^{-}&=\left(\mathcal{L}^{-}J_{0}^{-}+J_{1}^{-}\right)dx^{-}\,, \nonumber\\
    \hat{a}&=\left(\mathcal{L}\hat{J}_{0}+\hat{J}_{1}\right)d\phi\,,
\end{align}
where the Grassmann-valued $\psi$ is required to satisfy
\begin{equation}
   \partial_{-}\psi=0\,.\label{chiral2}
\end{equation}
For convenience, we now make the change $t=u$  and consider the following change of basis
\begin{equation}
J^{\pm}_{a}=\frac{\ell^{2}P_{a}\pm\ell Z_{a}}{2}\,,\qquad \hat{J}_{a}=J_{a}-\ell^{2}P_{a}\,, 
\end{equation}
\begin{equation}
 Q_{+}=\frac{\ell}{\sqrt{2}}\Tilde{Q}_{+},
\end{equation}
so that the asymptotic gauge field reads
\begin{eqnarray}
a &= & \left({\frac{1}{2}\cal M}du+\frac{1}{2}{\cal N}d\phi\right)Z_{0}+du\, Z_{1}+\left(\frac{1}{2}\mathcal{M}-\frac{1}{2\ell^{2}}\mathcal{F}\right)
d\phi\, J_{0}+d\phi\, J_{1}+\left({\frac{1}{2}\cal N}\,du+\frac{1}{2}\mathcal{F}d\phi\right)\,P_{0} \notag \\ 
& & +\frac{\Psi}{2^{1/4}}\Tilde{Q}_{+}d\phi\,+\frac{1}{\ell}\frac{\Psi}{2^{1/4}}\Tilde{Q}_{+}du\,,\label{aaa}
\end{eqnarray}
where we have redefined the arbitrary functions as follows
\begin{equation}
  \mathcal{M}=\left(\mathcal{L}^{+}+\mathcal{L}^{-} \right)\,, \qquad \mathcal{N}=\ell\left(\mathcal{L}^{+}-\mathcal{L}^{-} \right)\,, \qquad \mathcal{F}=\ell^{2}\left(\mathcal{L}^{+}+\mathcal{L}^{-}-2\mathcal{L}\right)\,,
\end{equation}
\begin{equation}
   \Psi=\frac{\ell}{\sqrt{2}}\psi \,.
\end{equation}
The conditions \eqref{chiral1} and \eqref{chiral2} now read
\begin{equation}
    \partial_{u}\mathcal{M}=\frac{1}{\ell^{2}}\partial_{\phi}\mathcal{N}\,, \qquad \partial_{u}\mathcal{N}=\partial_{\phi}\mathcal{M}\,, \qquad \partial_{u}\mathcal{F}=\partial_{\phi}\mathcal{N}\,,\label{con1}
\end{equation}
\begin{equation}
     \partial_{u}\Psi=\frac{1}{\ell}\partial_{\phi}\Psi\,.\label{con2}
\end{equation}
We can now apply the vanishing cosmological constant limit $\ell\rightarrow\infty$ directly to the $r$-independent asymptotic gauge field (\ref{aaa}), such that it reduces to
\begin{eqnarray}
a &= & \left({\frac{1}{2}\cal M}du+\frac{1}{2}{\cal N}d\phi\right)Z_{0}+du\, Z_{1}+\frac{1}{2}\mathcal{M}
d\phi\, J_{0}+d\phi\, J_{1}+\left({\frac{1}{2}\cal N}\,du+\frac{1}{2}\mathcal{F}d\phi\right)P_{0} \notag \\ 
& & +\frac{\Psi}{2^{1/4}}\Tilde{Q}_{+}d\phi\,\,,\label{agf}
\end{eqnarray}
which coincides with the asymptotic form of the gauge connection proposed in \eqref{aa}. As an ending remark, one can notice that these boundary conditions correspond to a supersymmetric extension of the ones proposed in \cite{Concha:2018zeb} for the Maxwell gravity theory, when the role of the generators $Z_a$ and $P_a$ is interchanged in \eqref{agf}. 

\section{Discussion}\label{concl}

In this paper, we have presented the CS supergravity theory based on the minimal supersymmetric extension of the Hietarinta symmetry. The CS theory can be seen as an extension of the $\mathcal{N}=1$ Poincaré CS supergravity theory \cite{Achucarro:1986uwr}. Although the Hietarinta superalgebra is isomorphic to the non-standard Maxwell superalgebra \cite{Lukierski:2010dy,Concha:2020atg}, the gauge field interpretation à la Hietarinta allows us the construction of a truly supergravity action. We have shown that the additional gauge field $\sigma^{a}$, which is responsible to turn on the torsion, modifies the asymptotic algebra. After considering appropriate boundary conditions, we showed that the asymptotic symmetry algebra of the bulk theory is given by an extension of the $\mathfrak{bms}_3$ superalgebra \cite{Barnich:2014cwa,Barnich:2015sca,Caroca:2019dds} with three non-trivial central charges. The obtained asymptotic symmetry algebra can alos be seen as a supersymmetric extension of the extended l-conformal Galiean algebra for $l=1$ \cite{Chernyavsky:2019hyp}. Interestingly, the extended $\mathfrak{bms}_{3}$ superalgebra can alternatively be recovered as a vanishing cosmological constant limit of three copies of the Virasoro algebra, one of which is augmented by supersymmetry. Let us mention that the analysis of the energy bounds and asymptotic Killing spinors is analogous to the one presented in \cite{Barnich:2014cwa} for three-dimensional Poincaré supergravity since the extended $\mathfrak{bms}_3$ superalgebra introduced here contains $\mathfrak{bms}_3$ as subalgebra. 

As a generalization of our results, it would be worth it to study the inclusion of a cosmological constant to the Hietarinta supergravity. In this direction, one expect that the symmetry algebra of the corresponding theory should be isomorphic to three copies of the $\mathfrak{so}\left(2,1\right)$ algebra, one of which should be augmented by supersymmetry. It would interesting to explore if its bosonic counterpart admits black holes solutions of the BTZ type \cite{Banados:1992wn,Banados:1992gq} that resemble the ones obtained in \cite{Aviles:2023igk} in presence of a non-vanishing torsion. A further more detailed analysis of the solutions and  thermodynamics of the Hietarinta gravity theory with and without cosmological constant might be of interest. 

It might also be of interest to explore higher-spin generalizations of the Hietarinta gravity model. In presence of spin-3, one could expect to find an extension of the Poincaré CS gravity coupled to spin-3 gauge fields \cite{Afshar:2013vka,Gonzalez:2013oaa,Gonzalez:2014tba,Matulich:2014hea}. A question is whether such spin-3 gravity model may appear as a flat limit of three copies of the $\mathfrak{sl}\left(3,\mathbb{R}\right)$ analogously to the spin-3 Maxwell gravity theory \cite{Caroca:2017izc}. At the asymptotic symmetry level, one could also expect to obtain an extension of the spin-3 version of the $\mathfrak{bms}_{3}$ algebra \cite{Afshar:2013vka,Gonzalez:2013oaa,Matulich:2014hea} which should be related to a combination of $\mathcal{W}_{3}$ algebra through a flat limit. A consistent coupling of the Hietarinta gravity theory with massless spin-$\frac{5}{2}$ gauge field might be also of interest. It would be interesting to verify if such higher-spin generalization extend the hypergravity of Aragone and Deser \cite{Aragone:1983sz} and the Poincaré CS hypergravity theory \cite{Fuentealba:2015jma,Fuentealba:2015wza}.

Another aspect that it would be worth exploring is the non-relativistic regime of the Hietarinta CS supergravity theory presented here.
Non-relativistic versions of supergravity theories have only been approached recently \cite{Andringa:2013mma,Bergshoeff:2015ija,Bergshoeff:2016lwr,Ozdemir:2019orp,deAzcarraga:2019mdn,Ozdemir:2019tby,Concha:2019mxx,Concha:2020tqx,Concha:2020eam,Concha:2021jos,Concha:2021llq,Ravera:2022buz,Bergshoeff:2022iyb,Bergshoeff:2023igy}. In this direction, the expansion method \cite{deAzcarraga:2002xi,deAzcarraga:2007et,Izaurieta:2006aj} has been a powerful tool to derive the corresponding non-relativistic counterpart of a three-dimensional supergravity theory \cite{deAzcarraga:2019mdn,Ozdemir:2019tby,Concha:2019mxx,Concha:2020tqx,Concha:2020eam,Concha:2021jos,Concha:2021llq}. Following the procedure used in \cite{Concha:2019mxx,Concha:2020tqx,Concha:2021jos,Concha:2021llq}, it would be interesting to obtain the corresponding non-relativistic Hietarinta supergravity [work in progress]. We guess that the non-Lorentzian Hietarinta superalgebra should contain the extended-Bargmann superalgebra \cite{Bergshoeff:2016lwr} as subalgebra. 

\section*{Acknowledgment}
This work was funded by the National Agency for Research and Development ANID - SIA grant No. SA77210097 and FONDECYT grant 1211077, 11220328 and 11220486. The authors would like to thank to Javier Matulich for enlightening discussions and comments.   The authors would also like to thank the Dirección de Investigación and Vice-rectoría de Investigación of the Universidad Católica de la Santísima Concepción, Chile, for their constant support. 
\section*{Appendix} \label{app}

\bibliographystyle{fullsort}
 
\bibliography{SuperHieta}

\end{document}